\documentclass[11pt]{article}
\usepackage{moriond,epsfig}

\def\Journal#1#2#3#4{{#1} {\bf #2}, #3 (#4)}

\def\NIMA{{\em Nucl. Instrum. Methods} A}

\def\PRL{\em Phys. Rev. Lett.}
\def\PRD{{\em Phys. Rev.} D}

\def\be{\begin{equation}}
\def\ee{\end{equation}}
\def\bea{\begin{eqnarray}}
\def\eea{\end{eqnarray}}

\newcommand{\anta}{{\sc Antares}}
\begin{document}
\vspace*{4cm}
\title{The \anta~Neutrino Telescope : \\
first results
}

\author{Thierry PRADIER \\
for the \anta~Collaboration \footnote{\texttt{http://antares.in2p3.fr}}
}

\address{University Louis-Pasteur \& Institut Pluridisciplinaire Hubert Curien \\
Subatomic Research Department (DRS) \\
23 rue du Loess BP 28 - F67037 Strasbourg, France}

\maketitle
\abstracts{The \anta~Collaboration is completing the deployment of a 12 lines underwater detector, 2500m deep in the Mediterranean Sea, dedicated to high energy neutrino astronomy. Starting with the first line in 2006, 10 lines were continuously recording data by the end of 2007, which allow us to reconstruct downward-going cosmic muons, and search for the first upward-going $\nu$-induced muons. Calibration topics will be described and preliminary results presented. 
}

\section{The neutrino as a new high-energy messenger}

The advantage of using neutrinos as new messengers lies firstly on their weak interaction cross-section ; 
unlike protons ($E_{\textrm{{\tiny cut-off}}}\sim 5\times 10^{19}$eV, $l_{\textrm{{\tiny free path}}}\sim 50$ Mpc) or $\gamma$ ($E_{\textrm{{\tiny cut-off}}}\sim 10^{14}$eV, $l_{\textrm{{\tiny free path}}}\sim 10$ Mpc), they provide a cosmological-range unaltered information from the very heart of their sources. 
Secondly, charged particles are deflected by magnetic fields, with a mean deflection $\Delta \theta \sim L(\textrm{kpc}) \frac{Z B(\mu G)}{E (\textrm{EeV})}$, yielding for Galactic Sources $\Delta \theta \sim 12^{\circ}$ at $10^{19}eV$. Neutrinos on the other hand point directly to their sources and exact production site.

The neutrinos \anta~is aiming at are typically TeV neutrinos from AGNs (supermassive black holes believed to be hosted in the center of each galaxy), typically 30 orders of magnitude lower in flux \cite{agn_flux} than solar neutrinos. The detection of those specific neutrinos requires under water/ice instruments, or alternatively acoustic/radio techniques in the PeV-EeV range and air showers arrays above 1 EeV.
In spite of efforts in those various energy ranges, since the detection of the MeV neutrino burst from SN 1987A by {\sc Kamiokande/Baksan/imb/Mont-Blanc} \cite{sn1987} no astrophysical source for neutrinos above a few GeV has ever been identified.

\section{TeV cosmic neutrinos : production and detection}

Sources for TeV $\nu$ are typically compact objects (neutron stars/black holes), from which often emerge relativistic plasma jets with a still unclear composition - leptonic or hadronic ?

\subsection{Sources of TeV cosmic neutrinos ?}

Most of these sources have already been extensively studied from radio wavelengths up to $\gamma$-rays. These photons can be produced by $e^-$ {\it via} inverse compton effect (on ambient photon field)/synchrotron radiation, or by protons/nuclei {\it via} photoproduction of $\pi^{0}/\pi^{\pm}$ :

\begin{equation}
\begin{array}{cccccc}
 p / A  +  p / \gamma & \longrightarrow & \pi^{0}~\pi^{\pm}, & \textrm{with~}\pi^{0} \longrightarrow \gamma \gamma, & \textrm{and~}\pi^{\pm} \longrightarrow \nu_{\mu}~\mu,  & \mu \longrightarrow \nu_{\mu} \nu_e e 
\end{array}
\label{eq:prodmu}
\end{equation}

\begin{figure}[h!]
\centerline{\includegraphics[width=0.9\linewidth]{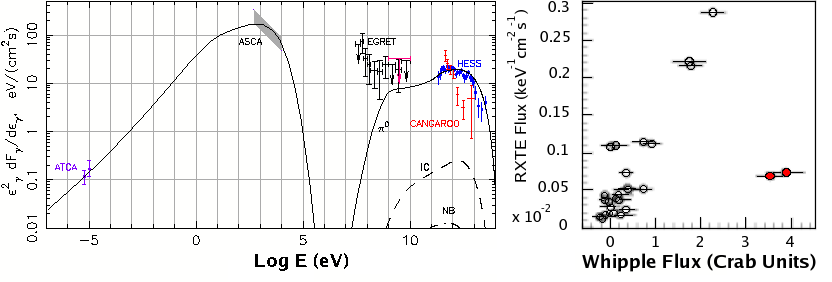}}
\caption{Left : Multiwavelength observations of the SNR RXJ 1713.7-39; the solid curve at energies above $10^7$~eV
  corresponds to $\pi^0$-decay $\gamma$-ray emission, whereas the dashed and dash-dotted
  curves indicate the inverse Compton (IC) and Nonthermal Bremsstrahlung (NB)
  emissions, respectively. Right : Whipple vs RXTE flux, for the Blazar 1ES1959+650, which shows the existence of orphan $\gamma$ flares (in red).
\label{fig:hints}}
\end{figure}

In the former scenario, no neutrinos are produced, whereas in the latter, the neutrino flux is directly related to the gamma flux: a TeV neutrino detection from gamma sources would then yield a unique way to probe the inner processes of the most powerful events in the universe. Several hints exist which indicates that hadrons could be accelerated up to very high energies. Firstly, the combined radio, X-rays and $\gamma$-rays observations of the shell-type supernova remnant RX J1713.7-3946 \cite{rx} favour the production of photons {\it via} $\pi^0$ decay (figure \ref{fig:hints}, left). Secondly, the correlations between X and $\gamma$ for the Blazar 1ES1959+650 \cite{1es} prove the existence of $\gamma$ flares not visible in X (figure \ref{fig:hints}, right), which is difficult to account for in purely leptonic models. Finally, it should be reminded that the so-called GZK cut-off (interaction of ultra-high energy cosmic rays with the CMB) is a guaranteed source of sub-EeV neutrinos \cite{gzk}.

\subsection{Practical issues for their detection}

\anta~can be seen as a fixed target experiment: a cosmic muon neutrino
interacts in the Earth and produces a muon that propagates in sea water. The \v{C}erenkov light emitted
by the muon is detected by an array of photomultipliers arranged in strings, able to reconstruct the energy and direction of
the incident muon/neutrino \cite{markov}.

The main physical backgrounds are twofold. Atmospheric muons ($\sim 1/s$ at the reconstruction level in \anta), produced in the upper atmosphere
by the interaction of cosmic rays, can be strongly suppressed because of their downward direction. Upward-going atmospheric neutrinos ($\sim$ 10/days in \anta) on the
other hand are
more delicate to identify: they have exactly the same signature as the expected cosmic signal \anta~awaits for.

For a given neutrino flux $\Phi_{\nu}$, the number of events expected for a telescope of effective area $A_{\mu}$ ({\it i.e.} the size of a detector 100$\%$ efficient for muons) can be estimated as follows:

\begin{equation}
N_{\mu} \propto \Phi_{\nu} \times P_{\textrm{absorption}}(\theta,E_{\nu}) \times \sigma_{\nu} \times R_{\mu} \times A_{\mu},
\label{eq:nmu}
\end{equation}

\noindent
where typically $\sigma_{\nu} \approx 2 \times 10^{-34}~cm^2$, $R_{\mu} \approx 10~km$, and $A_{\mu} \approx 0.06~km^2$ (roughly the geometrical surface for \anta~for reconstructed events with an angular resolution below 1$^{\circ}$) at 100 TeV.

\begin{figure}[h!]
\centerline{\includegraphics[width=0.9\linewidth]{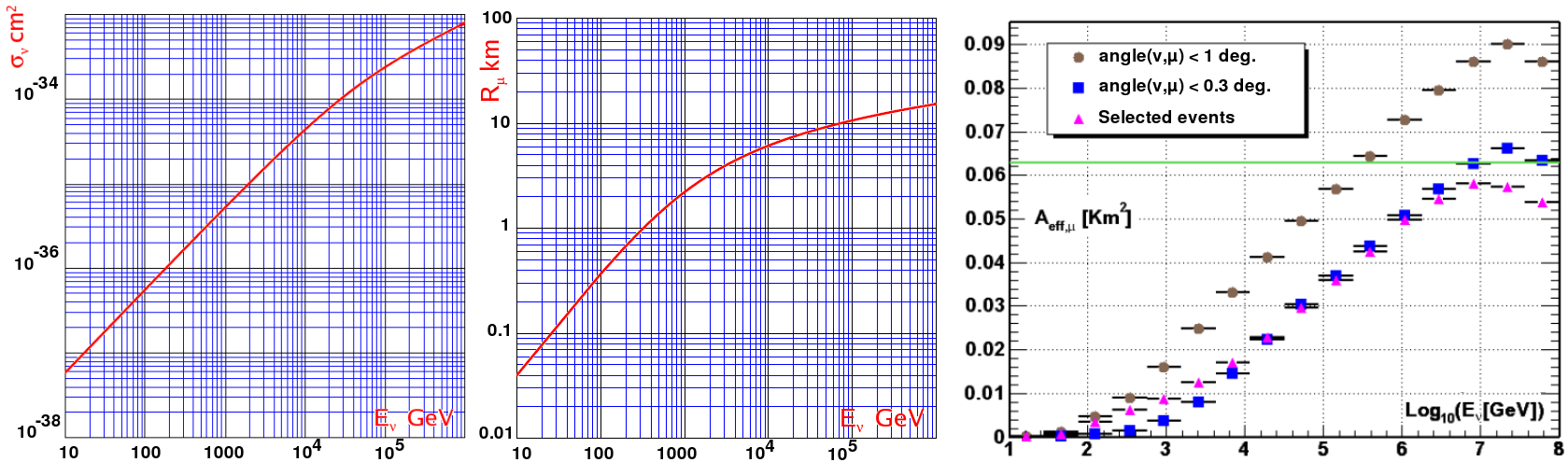}}
\caption{$\nu$ interaction cross-section, $\mu$ range in water, and effective area for muons.
\label{fig:murange}}
\end{figure}

The neutrino luminosity $L_{\nu} = 4\pi d^2 \Phi_{\nu}$ needed to detect $N_{\nu}$ events can then be written :

\begin{equation}
L_{\nu} \approx 10^{46} N_{\nu} \left( \frac{d}{4~\textrm{Gpc}}\right)^2 \left( \frac{E_{\nu}}{100~ TeV}\right)^{1-\alpha} \left( \frac{A_{\mu}~T_{\textrm{obs}}}{km^2~yr}\right)^{-1} \textrm{erg/s},
\label{eq:lnu}
\end{equation}

\noindent
for a source observed over a time $T_{\textrm{obs}}$ ; $\alpha \sim 1/0.5$ below/above 100 TeV. Typically for blazars ($d \sim \textrm{Gpc}$ and $L \sim 10^{47} \textrm{erg/s}$), the required effective area is $A_{\mu} \sim 1~km^2$, far beyond the reach of \anta. For galactic sources and $L \sim 10^{35} \textrm{erg/s}$, the necessary effective area goes down to $A_{\mu} \sim 0.1~km^2$, typically the size of \anta.

\section{\anta~: description, performances \& milestones}

Two main kinds of signals can be detected with \anta: $\mu$ tracks initiated by the charged current interaction of a $\nu_{\mu}$ in the Earth, and showers produced by the interaction of a neutrino (mainly $\nu_e$ and $\nu_{\tau}$ by charged or neutral current channels) in water. Those signals are faint signals, and because of light scattering and absorption in water, their detection require single-photoelectron-sensitive devices. The measurements of the time of the hits (time resolution of the order of ns) and the amplitude of the hits (with a resolution of about  $30 \%$), together with the position of the hits (by measuring the position of each PMT, to reach a resolution of about 10 cm) are needed to achieve the reconstruction of those signals with the desired resolution.

Muon tracks are detected {\it via} their directional \v{C}erenkov light (angle in water $\approx 42 ^{\circ}$) and can be reconstructed with an angular resolution below 0.3$^{\circ}$ above 10 TeV (the resolution below this energy is dominated by the kinematics of the interaction). The energy resolution is quite poor, a factor 2-3 on average, restricted by the granularity/density of the light sensors and the fact that the muon traverses the detector. Showers produced by $\nu_e$ on the other hand emit quasi-isotropic light, and can be reconstructed with a better energy resolution (roughly 30 \%) but with a poorer angular resolution~$\sim$~3-5~$^{\circ}$.

\subsection{Detector description}

\begin{figure}[h!]
\centerline{\includegraphics[width=0.5\linewidth]{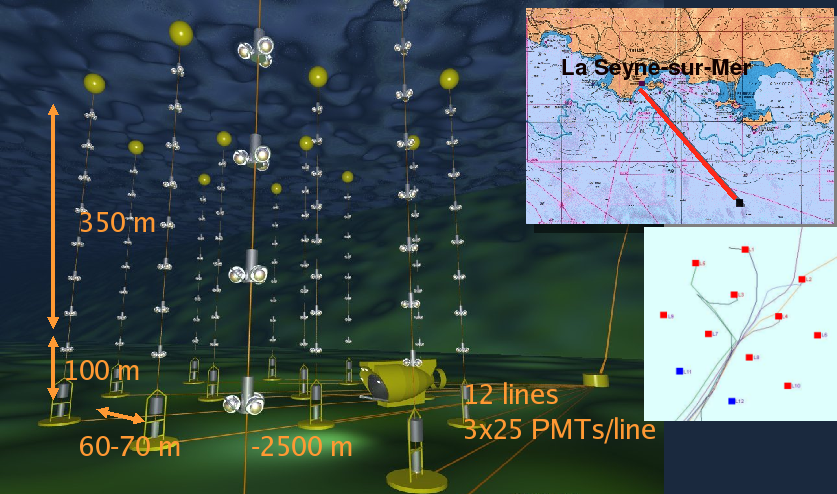}\includegraphics[width=0.5\linewidth]{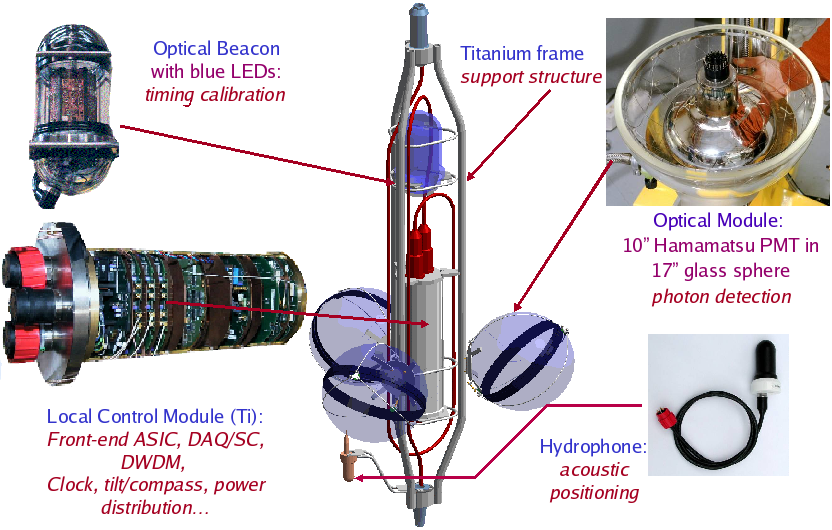}}
\caption{Left : Description, position and layout of the 12 lines of the \anta~telescope, 10 of which are currently taking data. Right : Instruments on board of one of the \anta~storeys.
\label{fig:layout_storey}}
\end{figure}

The \anta~neutrino telescope, deployed at 2500 m below sea surface, 40 km off the coast of Toulon (Southern France) is composed of 12 strings, with 25 storeys each containing a triplet of 10$^{``}$ photomultipliers oriented at 45 degrees downward to be optimally sensitive to upward going muons. As of March 2008, 10 lines were connected and continuously taking data since end of 2007: the first line was operating as soon as March 2006, the second line in September 2006, and in January 2007 5 lines in whole were operational. A schematic description of the detector, together with the layout of the lines, can be found in figure \ref{fig:layout_storey} (left plot). The full completion of the telescope should be performed by summer 2008.

An instrumented line is also present on site, to perform environmental measurements : sea water temperature, salinity, sound velocity probes, as well as speed of the sea current and direction, all parameters required for an optimum track reconstruction, and for studies of biological backgrounds.
The quality of sea water in particular, and its knowledge, is a fundamental parameter for \v{C}erenkov photon detection \cite{anta_sites}. The absorption length at the \anta~site at 470 nm is roughly 60 m, with an effective scattering length of 300 m. It is a combination of this water quality (scattering and chromatic dispersion, accounting for 1.5 ns at a distance of 40m) and of the timing performances described below which finally takes down the angular resolution at the 0.2$^{\circ}$ level at high energy (where the neutrino and the produced muon are essentially colinear).

The right panel of figure \ref{fig:layout_storey} displays the content of one of the \anta~storey. PMTs are enclosed in pressure-resistant spheres \cite{anta_pmt}, and a Titanium cylinder contains the front-end electronics. The intrinsic photoelectron transit time spread between the photocathode and its first dynode is roughly 1.3 ns, and the last dynode signal being digitised by a devoted chip, the ARS, gives a resolution better than 0.5 ns.
The tilt/compass cards, and hydrophones on some of the storeys, allow us to measure continuously the position of the optical modules (see section~\ref{sec:calib}). Finally, time calibration (see section \ref{sec:calib}) can be performed using a laser and LED beacons.


\subsection{Physics performances}

The energy resolution is a crucial element for the study of diffuse $\nu$ flux. The link between extra-galactic sources of both cosmic rays, $\gamma$-rays and $\nu$ leads to severe limits on the $\nu$ diffuse flux expressed in the Waxman-Bahcall (WB) upper bound \cite{wb_limit} $E^2\Phi<4.5\times10^{-8} \textrm{GeV}.\textrm{cm}^{-2}.\textrm{s}^{-1}.\textrm{sr}^{-1}$. After 3 years, \anta~is expected to set an upper limit of $E^2\Phi<3.9\times10^{-8} \textrm{GeV}.\textrm{cm}^{-2}.\textrm{s}^{-1}.\textrm{sr}^{-1}$, just below the WB estimate. 

\begin{figure}[h!]
\centerline{\includegraphics[height=6cm]{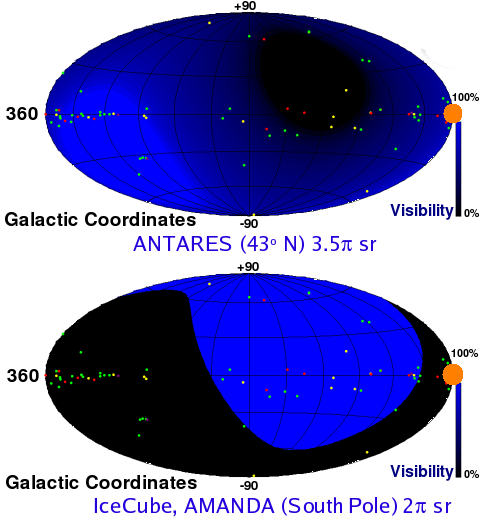}\includegraphics[height=6cm]{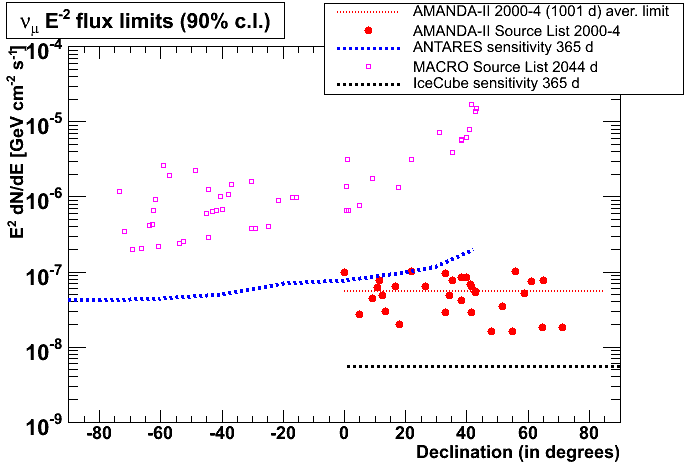}}
\caption{Left : Sky as visible by \anta~and {\sc Amanda/IceCube}, in Galactic coordinates, the circle indicating the Galactic Centre. Right : Sensitivity for Point-like sources.
\label{fig:sky_nulimits}}
\end{figure}

The \anta~sensitivity to point-like sources can be estimated as a function of the declination of a potential source: figure \ref{fig:sky_nulimits} (left) shows that \anta~will be able to observe the Galactic Centre and other interesting $\gamma$ sources, for most of the time complementary to {\sc IceCube}. The $90\%$ upper limit for $\nu_{\mu}+\bar{\nu}_{\mu}$ flux in case of null signal after 1 year is $E^2\frac{dN}{dE_{\nu}} = 4\times10^{-8} \textrm{GeV}.\textrm{cm}^{-2}.\textrm{s}^{-1}$ at $\delta=-90^{\circ}$, and rises to $1.5\times10^{-7} \textrm{GeV}.\textrm{cm}^{-2}.\textrm{s}^{-1}$ at $\delta=+40^{\circ}$. Those limits improve those of {\sc macro} for the Southern Sky, as can be seen in figure \ref{fig:sky_nulimits} (right), and are comparable to those obtained by {\sc amanda II} for the Northern Sky \cite{limits}.

\subsection{\anta~Milestones}

Conceiving and building a Neutrino Telescope in the Mediterranean require much preparatory work: the proposal of the experiment dates back to 1999 \cite{anta_proposal}. Nine years were thus needed to realise the 10 lines (soon 12!) that are currently taking data. Here is a short historical overview, before describing the calibration and results of \anta:

\begin{itemize}
\item {\bf 1996-2000 : Validation of the Project - } Water properties were first studied in order to choose the best site, and marine technologies were developed and improved. This period ended with the deployment of a demonstrator line \cite{anta_line5} and the reconstruction of the first atmospheric muons in the \anta~Collaboration.

\item {\bf 2001-2004 : Final {\sc r\&d}, first deployments - } The Electro-Optical cable between the shore and the site was deployed in 2001, the Junction Box (distribution of power to lines) was operational in 2002. Finally, a Prototype Sector Line (similar to a final \anta~line, but with only one sector consisting of 5 storeys) sucessfully took data between end of 2002 until its recovery in July 2003 ; a Mini-Instrumentation Line ({\sc mil}, environmental probes mainly) was also operated for a few months between Feb. and May 2003 \cite{anta_psl}.

\item {\bf 2005-2007 : Construction, deployment and operation - } The {\sc mil} was recovered to be upgraded with two storeys of Optical Modules ({\sc milom}), and took data for 2 years (April 2005 - March 2007), before the deployment, connection and operation of the first complete \anta~line in March 2006 \cite{anta_milom}.

\end{itemize}

\section{\anta~in operation : calibration of a neutrino telescope}
\label{sec:calib}

To be able to extract physical results from raw data, a neutrino telescope like any other detector has to be understood and calibrated : some aspects of this calibration will be reviewed here.

\subsection{Acoustic positionning}

\begin{figure}[h!]
\centerline{\includegraphics[width=0.5\linewidth]{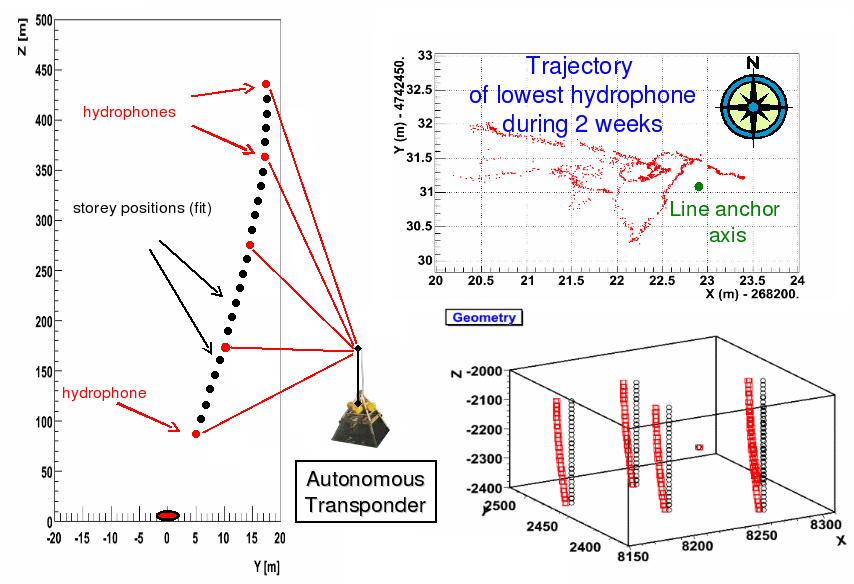}\includegraphics[width=0.5\linewidth]{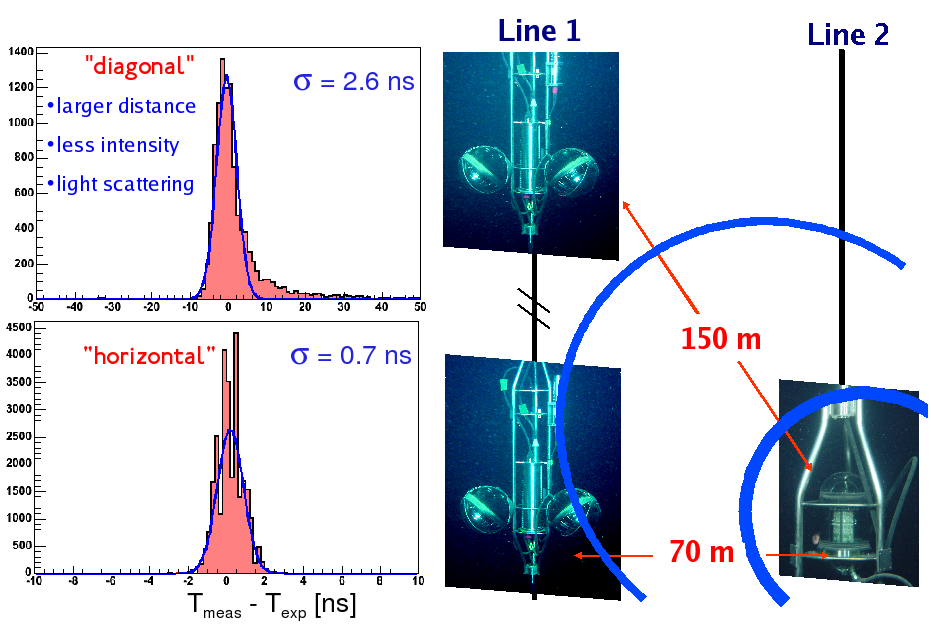}}
\caption{Left : Principles of acoustic positioning. Right : Principles of Time Calibration with LED Beacons.
\label{fig:acoustic_timecalib}}
\end{figure}

The particularity of an underwater neutrino telescope, as compared with a $\nu$ Telescope in ice ({\sc IceCube}), is that the lines, maintained as vertical as possible with a buoy, are moving under the influence of water currents. Hence, a reconstruction of  the line shape is needed for the positions of each individual PMTs to be known with an accuracy of 10-cm, which is required to achieve the 0.2$^{\circ}$ angular resolution at high energy. This is performed by an acoustic positioning system~\cite{anta_acoustic}, as shown in figure \ref{fig:acoustic_timecalib} (left panel). Transponders on the sea ground emit signals, detected by hydrophones equipping some of the storeys. Together with data from tilt/compass cards, this allows for the determination of shape of the line; the actual position of the top storey can differ from the original straight line position by up to 15 m for strong sea currents ! If not accounted for, this would imply an error in the absolute positioning of a source of several degrees. 

\subsection{Time calibration}

An error of 0.3 ns in the photon arrival time on one of the PMTs is equivalent to a 10-cm error on its position: timing performances are thus as primordial as the line positioning. They can be studied using the light emitted by LED Beacons \cite{anta_led}. This blue light is detected by PMTs on adjacent lines, and the timing resolution as shown in figure \ref{fig:acoustic_timecalib} (right panel) can be estimated to be ({\it horizontal} case) as low as 0.7~ns, which is then dominated by the electronics.

\subsection{$^{40}K$ calibration}

Sea water contains $^{40}$K which is a $\beta$ emitter, the $e^-$ in turn emitting \v{C}erenkov radiation. Adjacent PMTs can thus coincidently detect this light, and this $^{40}K$ calibration is a powerful way to estimate the acceptance of each optical module \cite{anta_k40}.

\section{\anta~in operation : first signals and selected results}

\begin{figure}[h!]
\centerline{\includegraphics[width=0.8\linewidth]{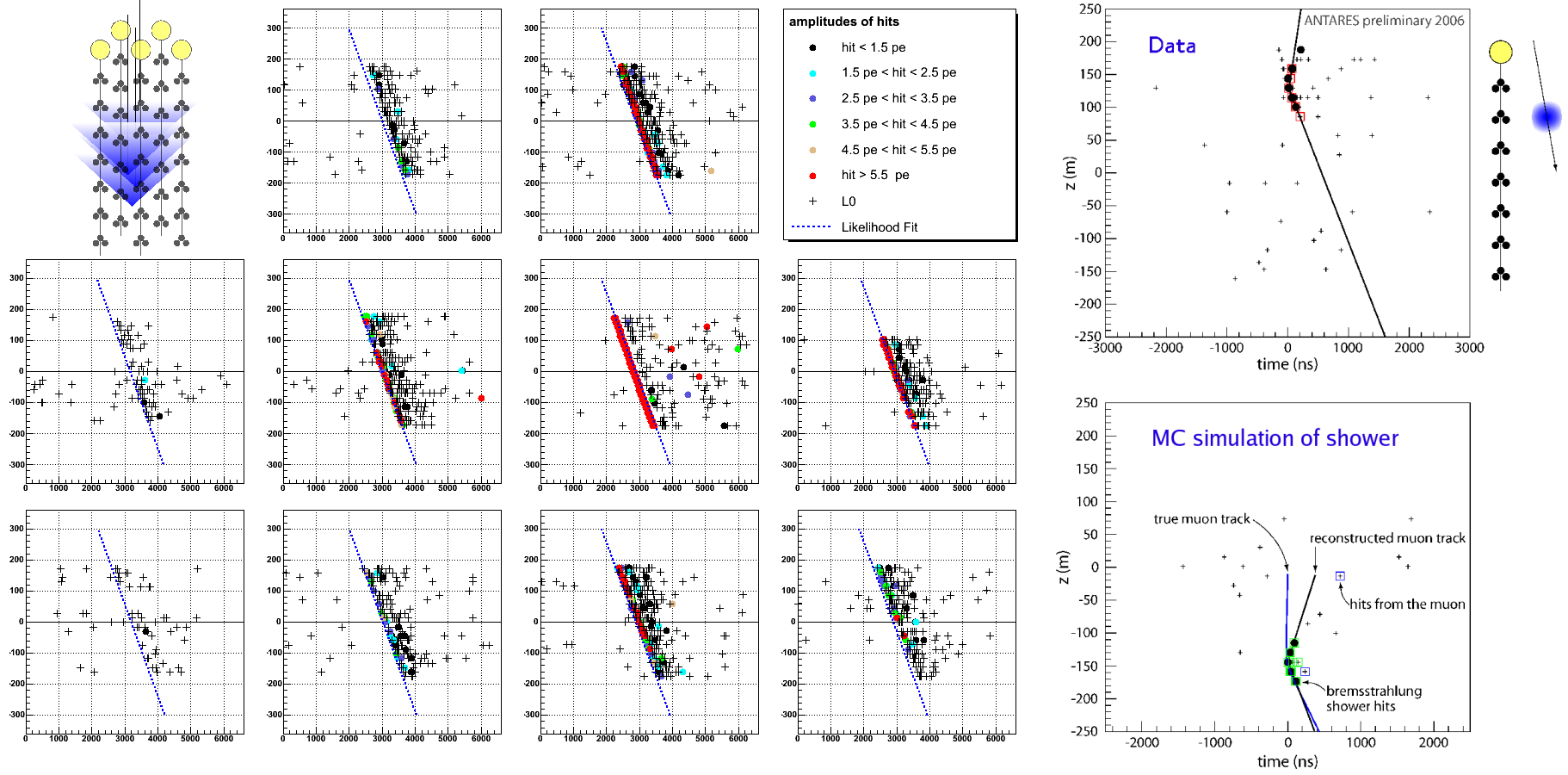}}
\caption{A $\mu$ bundle seen in 10 lines (altitude of hits {\it vs} time of hits, left) and a shower in 1 line (right), from Line 1 data (top) and Monte-Carlo (bottom).
\label{fig:bundle_shower}}
\end{figure}

The trigger rate of \anta~is roughly 1/s, mostly corresponding to atmospheric muons, 70\% of which are multiple quasi-parallel muons, arriving at the same time in the detector \cite{anta_bundles}. A nice muon bundle seen with 10 lines is displayed in figure \ref{fig:bundle_shower}. Showers developping along a $\mu$ track can also be observed (fig. \ref{fig:bundle_shower}, right).

\subsection{Line 1 data : first estimate for atmospheric muons flux}

The angular distribution of reconstructed events can be transformed into an intensity 
{\it versus} depth (using acceptance corrections from simulation) in the region of uniform acceptance: each value of the zenith angle corresponds to a certain slant depth through the water mass above the detector. To compute the muon vertical intensity, the distribution of muons at sea level has to be taken into account \cite{anta_line1}. The results obtained using Line 1 data, with low sea currents from May to September 2006 (equivalent live time 10 days), are shown in figure \ref{fig:thetanu_line1muflux}. The errors of the order of 50\% are dominated by the PMT acceptance. The agreement between data and other published values is good, showing that physics results can be extracted even with only 1 line!

\subsection{Data with 5 lines : neutrino candidates}

\begin{figure}[h!]
\centerline{\includegraphics[width=\linewidth]{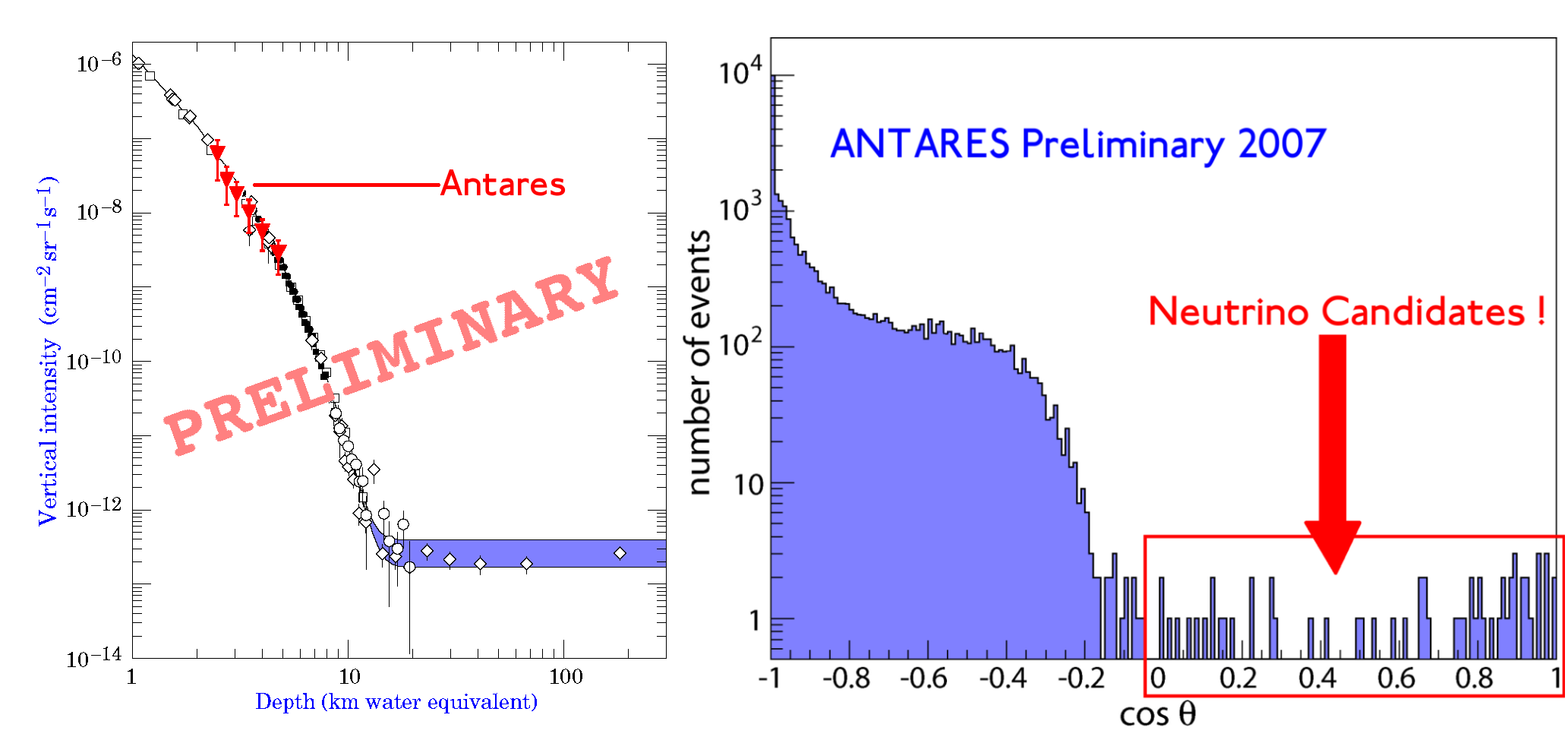}}
\caption{Left : Line 1 data, vertical intensity of atmospheric muons {\it versus} depth (water equivalent). Right : 5 lines data, distribution in zenith angle of (selected) events. 
\label{fig:thetanu_line1muflux}}
\end{figure}

Figure \ref{fig:thetanu_line1muflux} (right) shows a zenith angle distribution obtained with 5 Lines data (February-May 2007, equivalent live time 54 days). These data contain roughly $5\times10^6$ events, reconstructed with a $90 \%$ efficiency. After quality cuts, 20000 events remain, for which $\cos\theta$ is shown. The events reconstructed as upgoing ($\cos\theta > 0.1$) are 55 neutrino candidates, the events reconstructed as downgoing corresponding to atmospheric muons. The peak at -1 are vertically downward-going atmospheric $\mu$, or muon bundles, very nicely reconstructed. Finally the slight excess near $\cos\theta \sim 1$ is an acceptance effect : the telescope is more sensitive to purely vertical tracks. 

\newpage
\section{Conclusions and outlook}

The 2 remaining \anta~lines should be taking data by summer 2008, giving birth to the biggest Neutrino Telescope in the Northern Hemisphere. The current 10-lines telescope is already operating, and its acoustic positioning  is fully functional. Despite its smaller size with respect to {\sc IceCube} \cite{icecube}, \anta~observes the Galactic Centre and other potential sources of TeV $\nu$ not accessible from the South Pole, leaving some margins for unexpected discoveries.

Furthermore, \anta~is a part of the {\sc gcn}, Gamma-ray bursts Coordinates Network \cite{gcn}, dedicated to $\gamma$-ray bursts, thought to be potential sources of high energy neutrinos : satellites/telescopes broadcast real-time alerts, which in turn trigger the recording of all \anta~data within a 2 minutes time window \cite{anta_gcn}. Over a period of 15 months, 172 {\sc gcn} alerts were distributed to \anta, and the telescope took data for 152 of them, corresponding to a $\sim 90\%$ live time !
%

\anta~must be seen as the first stage towards a km$^3$-scale telescope, for which European institutes involved in current $\nu$ astronomy projects (\anta, {\sc Nemo}, {\sc Nestor}) are already collaborating. This network, {\sc KM3NeT} \cite{km3net}, will give birth to a telescope with which neutrinos will be as common messengers as gamma-rays are now.

\end{document}